# The Magnetic Origin of $d_{x^2-y^2}$ High-$T_c$ Superconductivity from Tunneling Spectroscopy Measurements on $Bi_2Sr_2CaCu_2O_{8+x}$ Single Crystals.

Andrei Mourachkine

*Université Libre de Bruxelles, Service de Physique des Solides, CP233, Boulevard du Triomphe, B-1050 Brussels, Belgium*

**Abstract.** The complete scenario of high-$T_c$ superconductivity based on experimental data from electron-tunneling spectroscopy on underdoped, overdoped and Ni-doped $Bi_2Sr_2CaCu_2O_{8+x}$ single crystals using a break-junction technique is presented. There are two different types of superconductivity in cuprates: superconductivity due to pairing of spinons on charged stripes and magnetic superconductivity mediated by spin-waves. The coherent state of the spinon superconductivity is established via the magnetic superconductivity. In $Nd_{2-x}Ce_xCuO_4$, there is only the spinon superconductivity and the coherent state is established due to the Josephson coupling between charged stripes. Below $T_c$, we observe four different gaps, namely, (i) a SDW gap due to antiferromagnetic correlations; (ii) a superconducting gap due to spinon pairing; (iii) a d-wave magnetic superconducting gap, and (iv) a small superconducting gap most likely having g-wave symmetry. We show that the d-wave superconductivity is mediated by spin-waves and the magnitude of the superconducting gap due to pairing of spinons is larger than the magnitude of the d-wave gap, however, the d-wave gap is more intense. The superconducting gap due to pairing of spinons most likely has a s-wave symmetry. The maximum magnitudes of the SDW, the spinon and d-wave gaps are located at ($\pi/2$, $\pi/2$), ($\pi/2$, $\pi/2$) and ($\pi$, 0) on the Fermi surface, respectively. The d-g-wave superconductivity mediated by spin-waves can be considered as a pairing of magnetic polarons. We formulate a theorem for cuprates by analogy with the Anderson's theorem for classical superconductors. The presented model of high-$T_c$ superconductivity naturally explains other experimental data.

**PACS.** 74.20.Mn Nonconventional mechanisms - 74.50.+r Tunneling phenomena - 74.72.Hs Bi-based cuprates

## 1 Introduction

Since the discovery of high-temperature superconductors (HTSC) in 1986 by Bednorz and Müller [1] the mechanism of superconductivity (SC) in copper-oxide materials remains an open question. The HTSC are among the most complex systems studied in condensed-matter physics. Although there has been no agreement on a complete theory, a consensus has formed that the strong electron repulsion and quasi-two-dimensional layered nature of these materials are responsible for their anomalous physical properties and high $T_c$. Many experiments show some universality of physical properties across widely different materials with different values of $T_c$. The definitive confirmation of the predominant $d_{x^2-y^2}$ (hereafter, d-wave) character of the SC in the cuprates is a great advance in recent years. However, the interpretation of some experimental data suggests the presence of a s-wave component. The existence of a pseudogap (PS gap) in electronic excitation spectra of the HTSC, which appears below a certain temperature $T^* > T_c$, is considered to be amongst the most important features of cuprates. There is a clear microscopic difference between the classic BCS superconductors and the HTSC, namely that they have a different origin and that different criteria are required for the HTSC than for the classic SCs. As for the bulk characteristics, the controlling factor in cuprates seems only to be the hole density in $CuO_2$ planes. The parent compounds of the HTSC are antiferromagnetic (AF) insulators with a large value of the exchange constant, $J = 0.15$ $eV$. This large value of $J$ is considered to be responsible for the high $T_c$. At the same time, there are some differences among different families of cuprates, the most remarkable one is a s-wave symmetry of the order parameter (OP) in $Nd_{2-x}Ce_xCuO_4$ (NCCO).

Magnetic polarons for the first time have been observed by Tsui, Dietz and Walker [2]. They reported the observation of inelastic spin-wave excitations in an antiferromagnetic $NiO_2$ thin layer by electron tunneling at 1 K. They found that (i) each tunneling electron is dressed on the average by two virtual magnons; (ii) the electron-magnon interaction is strong, and (iii) the electron-phonon interaction in $NiO_2$ is weak in comparisons with electron-magnon interaction. The observed conductance peaks corresponding to one, two, and three magnon excitations are not periodically spaced but the distance between the peaks decreases with increase of the number of excited magnons. This fact indicates that the magnon-magnon interaction is strong. They found that the AF $NiO_2$ barrier is considerably lower for tunneling electrons than that of the simple metal oxides.

There is clear evidence for charged stripe formation in $La_{1.6-x}Nd_{0.4}Sr_xCuO_4$ [3] and $YBa_2Cu_3O_{6.6}$ (YBCO) [4]. Holes induced in $CuO_2$ planes segregate into periodically-spaced stripes that separate AF domains. The period of the charge modulation is half of the spins. The density of holes in the charged domain walls is 0.5 hole per Cu site. The existence of a striped phase in doped 2D antiferromagnets has also been a subject of intense theoretical investigations. Emery, Kivelson and Zachar [5] have presented a theory of spin-gap proximity effect mechanism of underdoped HTSC based on charged stripe formation. They emphasize that any theory involving real-space pairs (bosons) tightly bound by a field of a pure electronic origin are *a priori* implausible due to the strong short-range Coulomb repulsion between two carriers.

Recent inelastic neutron scattering (INS) experiments have shown the presence of sharp magnetic collective mode ("resonance peak") in the SC state of YBCO [6,7]. As doping decreases the peak frequency decreases. The resonance peak shifts to higher frequencies with decreasing temperature [7,6]. This "collective mode" could partially be explained [6] by the spin fluctuation theory which is developed in the studies of Monthoux, Pines, Scalapino, Schrieffer, Sushkov, and others supporting the idea of the magnetic fluctuations mechanism of pairing [6,8-16]. Zaanen

[17], White and Scalapino [16] showed that the striped phase can coexist with the SC mediated by spin-waves.

The tunneling spectroscopy has played a crucial role in the verification of the BCS theory [18] but up to now the leading role in our understanding of the physics of the HTSC belonged to angle-resolved photoemission (ARPES) measurements [19,20]. The tunneling spectroscopy is a powerful tool. The tunneling spectroscopy is particularly sensitive to the density of state (DOS) near the Fermi level ($E_F$) and, thus, is capable of detecting *any* gap in the quasiparticle excitation spectrum at $E_F$ [18]. In addition to this, it has a very high energy resolution (less than $k_B T$ for a superconductor-insulator-superconductor (SIS) junction) [18]. The tunneling spectroscopy performed by a STM and a break-junction (B-J) technique has an additional advantage: to measure the DOS *locally*. The B-J technique can detect the Josephson current, hence, *can distinguish* between a superconducting and non-superconducting order parameter (OP).

The structure of the paper is as follows. Experimental details are described in Section 2. In Section 3, we consider the pseudogap above $T_c$. In Section 4, we focus attention on non-SC gaps in tunneling spectra below $T_c$. In Section 5, we concentrate on the origins of the non-SC gaps in tunneling spectra. The d-wave superconducting gap is discussed in Section 6. In Section 7, we examine the sub-gap structure in tunneling spectra. In Section 8, we consider the relations among different OPs. The model of the HTSC is presented in Section 9. In the context of the presented model, we discuss some important issues regarding the HTSC in Section 10. The Anderson's theorem for cuprates is formulated in Section 11. The final conclusions are presented in Section 12.

## 2 Experimental

The single crystals of Bi2212 were grown using a self-flux method and then mechanically separated from the flux in $Al_2O_3$ or $ZrO_2$ crucibles [21]. The dimensions of the samples are typically $3 \times 1 \times 0.1$ mm$^3$. The chemical composition of the Bi-2212 phase corresponds to the formula $Bi_2Sr_{1.9}CaCu_{1.8}O_{8+x}$ in overdoped crystals as measured by energy dispersive X-ray fluorescence (EDAX). The crystallographic *a, b, c* values of the overdoped single crystals are of 5.41 Å, 5.50 Å and 30.81 Å, respectively. The $T_c$ value was determined by either dc-magnetization or by four-contacts method yielding $T_c = 87 - 90$ K with the transition width $\Delta T_c \sim 1$ K. Some overdoped single crystals were carefully checked out to ensure that they are in an overdoped phase: the $T_c$ value was increasing up to 95 K when some oxygen was chemically taken off the samples. Underdoped samples were obtained from the overdoped single crystals by annealing the crystals in vacuum [22]. The underdoped samples which were studied in detail have the critical temperatures of 21 K (one sample), 51 K and 75 K.

The single crystals of Ni-doped Bi2212 were grown using also the self-flux method. The chemical composition of the Bi2212 phase with $T_c$ = 75 - 76 K corresponds to the formula $Bi_2Sr_{1.95}Ca_{0.95}(CuNi)_{2.05}O_{8+x}$ as measured by EDAX. The content of Ni is about 1.5 % with respect to Cu.

Experimental details of our B-J technique can be found elsewhere [23]. Here we present a short description of some technical details. Many break junctions were prepared by gluing a sample with epoxy on a flexible insulating substrate and then were broken by bending the substrate with a differential screw at 14 - 18 K in a helium atmosphere. By changing the distance between two pieces of a single crystal by a differential screw, it is possible to obtain a few tunneling spectra in one B-J. The normal resistance ($R_N$) of break junctions outside of the gap ranged from 50 Ω to 50 MΩ. The tunneling current-voltage characteristics $I(V)$ and the conductance curves $dI/dV(V)$ were determined by the four-terminal method using a standard lock-in modulation technique. The electrical contacts (typically with a resistance of a few Ω) were made by attaching gold wires with silver paint. The sample resistance (with the contacts) at room temperature varied from 10 Ω to about 2 kΩ, depending on the sample.

Typical conductance curves $dI/dV(V)$ and current-voltage $I(V)$ characteristics for our SIS break junctions on Bi2212 single crystals can be found elsewhere [23,24]. They exhibit the characteristic features of typical SIS junctions [25,26]. The magnitude of the SC gap can, in fact, be derived directly from the tunneling spectrum. However, in the absence of a generally accepted model for the gap function and the DOS in HTSC, such a quantitative analysis is not straightforward [27]. Thus, in order to compare different spectra, we calculate the gap amplitude $2\Delta$ (in m$e$V) as a half spacing between the conductance peaks at $\pm 2\Delta$.

## 3 The pseudogap

Most of our study of the PS gap was carried out on slightly overdoped Bi2212 single crystals [22]. Figure 1 shows a set of normalized tunneling conductance curves measured between 14 K and 290 K as a function of bias voltage on an overdoped Bi2212 single crystal with $T_c$ = 88.5 K. The value of the SC gap at 14 K is $2\Delta$ = 45 m$e$V. All curves, except the last one, show a gaplike structure. There is no sign indicating at what temperature the SC gap was closed. Across $T_c$, the SC tunneling spectra evolve continuously into a normal state quasiparticle gap structure which vanishes at 232 K < $T^*$ < 290 K but remains almost unchanged with temperature up to 232 K.

Figure 2 shows tunneling spectra measured on another overdoped Bi2212 single crystal with $T_c$ = 88 K. To our knowledge, the data shown in Fig. 2 are the most detailed data of the PS gap presented in the literature. The spectra in Fig. 2 are asymmetrical about zero bias. Such behavior is typical for tunneling spectra obtained on the nonmetallic surfaces [28]. The spectra at negative bias in Fig. 2 look similar to the tunneling spectra above $T_c$ shown in Fig. 1. It is clear from Fig. 2 that $T^* \approx$ 280 - 290 K. However, just above $T_c$, there is a gap-like structure which disappears at temperature $T_2^*$ lower than $T_1^* = 285$ K. In order to have an idea about the value of the $T_2^*$, in Figs. 3 and 4 we present the odd conductance $G_o(V) \equiv [G(V) - G(-V)]/2$ [2] for the tunneling spectra shown in Fig. 2 and the temperature dependence of the $G_o(V)$, respectively. One can see in Fig. 4 that the value of the $T_2^*$ is of the order of 130 K.

In general, the PS gap may be a charge-density-wave (CDW) gap, a spin-gap due to AF correlations (or a spin-density-wave (SDW) gap) or precursor pairing, or their combination. We will discuss possible models for the PS gap further.

## 4 The non-superconducting gaps below $T_c$

Figure 5 shows a set of tunneling spectra measured at 14 K on an overdoped Bi2212 single crystal with $T_c = 89.5$ K. The spectra B, C, D and E were measured in sequence by changing the distance in a junction (by bending the flexible substrate). The variations of the magnitude of the tunneling gap between 23 and 32.5 meV are in an excellent agreement with angle-resolved tunneling data [29] which are presented in Fig. 6. By comparing our data with the data presented in Fig. 6 we infer that the maximum of the tunneling gap (spectrum A in Fig. 5) is located at $(\pi/2, \pi/2)$ on the Fermi surface. The minimum of the tunneling gap (spectrum F in Fig. 5) is located at $(\pi, 0)$. By other words, Figure 5 presents the angle dependence of the tunneling gap between $(\pi/2, \pi/2)$ and $(0, \pi)$. We will analyze the spectra shown in Fig. 5 further. We concentrate now only on the Josephson current. One can see in Fig. 5 that the value of the Josephson current depends on the magnitude of the tunneling gap. It is noteworthy that such dependence is *typical* for each separate sample. The maximum values are slightly different for different samples.

The spectrum A in Fig. 5 has no Josephson current. However, the spectrum B in Fig. 5 already has a small Josephson peak. It is possible to explain the absence of the Josephson current in the spectrum A in Fig. 5 because the normal resistance of the junction is too high. However, in an underdoped Bi2212 single crystal we detected a weak Josephson current in a junction with $R_N \sim$ 1 MΩ. So, this is not the reason for the absence of the Josephson current. It must be underlined once more that such dependence of the value of the Josephson current on the magnitude of the tunneling gap is *typical*. Consequently, the spectrum with the maximum magnitude of the tunneling gap has, probably, a non-SC origin. Hereafter, we will call this tunneling gap as the excitation gap. The location of the maximum of the excitation gap is at $(\pi/2, \pi/2)$ on the Fermi surface. The $(\pi/2, \pi/2)$ direction on the Fermi surface coincides with the line connecting the Cu atoms (spins) on diagonal sites. Hence, it is most likely that the excitation gap has the magnetic origin.

Figure 7 shows tunneling spectra A and B measured at 14 K on an underdoped Bi2212 single crystals with $T_c$ = 51 K. The spectrum B corresponds to a large non-SC gap having at low bias a small piece of the curve corresponding to the combination of the SC and excitation gaps, which is very similar to the spectrum A at low bias. The difference between the spectra A and B corresponds to the DOS which is the combination of the SC DOS and the DOS corresponding the excitation gap. Some parts of the curve A - B in Fig. 7 are below zero because the spectra A and B were measured at slightly different angles on the Fermi surface. The shape of the conductance curve A - B (with the exception of the Josephson current) is very similar to the shape of a polaron (or bipolaron) [30]. Hence, the quasiparticles in the HTSC are *excitations*, not a real-space particles.

To our knowledge, this is for the first time a large non-SC gap which was detected below $T_c$ is presented in the literature. From Fig. 7, the origin of dips and humps in tunneling spectra outside the gap structure [25] are obvious. The humps correspond to the large non-SC gap which is shown in Fig. 7 (spectrum B). The dips appear naturally from the superposition of the large non-SC gap shown in Fig. 7 (spectrum B) and the excitation gap (or the SC gap which is the bound state of excitations). The large non-SC gap can be considered as a PS gap below $T_c$.

Let's do a small exercise. It was not possible below $T_c$ to detect separately the large non-SC gap in overdoped Bi2212. We will carry out the same procedure as we did in Fig. 7 for a spectrum measured on an overdoped Bi2212 single crystal, but we will use a created non-SC gap. Figure 8 shows the measured spectrum A on an overdoped Bi2212 sample with $T_c$ = 88 K. The shape of the spectrum A in Fig. 8 is typical for a maximum value of the tunneling gap in overdoped Bi2212 (see also the spectrum A in Fig. 5). The spectrum has no Josephson current and the minimum of the curve at zero bias is flat. The flat minimum is a part of the large non-SC gap. Thus, we have enough information to construct the large non-SC gap (curve B). The difference (curve A - B) between the measured spectrum A and constructed non-SC gap (curve B) corresponds to the pure DOS of the excitation gap (without of the SC DOS) in the overdoped Bi2212 sample. The shape of the constructed large non-SC gap is not very important, it would be even enough to draw straight lines between the dips and the flat minimum, the shape of the main tunneling peaks will be not affected drastically. The most important to note that the shapes of the curves A - B shown in Figs. 7 and 8 are similar and correspond to the DOSs of excitations.

Let's shortly summarize the data which we have on this stage. In Bi2212, above $T_c$, there is a PS gap. Below $T_c$, there are: (i) at least, one SC gap; (ii) an excitation gap, and (iii) a large non-SC gap (or a PS gap). The SC DOS and the DOS corresponding to the excitation gap are the DOSs of excitations (curves A - B in Figs. 7 and 8).

# 5 The origins of the non-SC gaps

Let's, first of all, analyze the origins of the excitation gap and the large non-SC gap which were detected below $T_c$. DeWilde, Miyakawa *et at*. [25] have performed tunneling measurements on the same set of optimally doped Bi2212 single crystals by STM [DeWilde] and by B-J technique [Miyakawa]. Main conductance peaks of any normal-state gap will have the same bias positions in a superconductor-insulator-normal (SIN) junction and in a SIS junction. The absolute values of bias of the location of conductance peaks corresponding to any SC gap will be twice larger in a SIS junction than in a SIN junction. In SIN junctions, the hump, the dip and main peak corresponding to the maximum value of the tunneling gap (*i. e.* to the excitation gap) have been detected at -150 meV, -80 meV and -37 meV [25], respectively. In SIS junction, they have been detected at 150 meV, 118 meV and 76 meV [25], respectively. There is no sense to discuss the location of the dips in tunneling spectra because they appear naturally from the superposition of the large non-SC gap and the excitation gap. So, it is clear that the large non-SC gap is a normal-state gap (150 meV = 150 meV). However, the excitation gap behaves like a SC gap (because 2×37 meV ≈ 76 meV). It is *remarkable* result that the excitation gap behaves like a SC gap and there is no Josephson current in tunneling spectra corresponding to this gap (see spectra A in Figs. 5 and 8).

The location of the maximum of the excitation gap is at ($\pi/2$, $\pi/2$) on the Fermi surface. The ($\pi/2$, $\pi/2$) direction on the Fermi surface coincides with the line connecting the Cu atoms (spins) on diagonal sites. One can see in Fig. 5 that the humps are much more pronounced in tunneling spectra when the tunneling gap is large, *i. e.* near ($\pi/2$, $\pi/2$) and almost absent in spectrum F in Fig. 5 which corresponds to the minimum of the tunneling gap, *i. e.* at ($\pi$, 0). This implies that the large non-SC gap is anisotropic with the location of the maximum at ($\pi/2$, $\pi/2$) on the Fermi surface. Tranquada *et al*. [3] have shown that charged stripes in cuprates develop along Cu-O bounds, *i. e.* in ($\pi$, 0) and (0, $\pi$) directions on the Fermi surface. Consequently, if we assume that the large non-SC gap is a CDW gap, it is simply impossible that the maximum of this gap is located in ($\pi$, $\pi$) direction. The ($\pi$, $\pi$) direction corresponds to the maximum of spin-spin interactions in cuprates. Hence, we infer that the large non-SC gap is a spin-gap due to AF correlations. Hereafter, we will call the large non-SC gap as the SDW gap. The presence of a SDW order below $T_c$ implies it's presence above $T_c$. We will show in the next Section that a CDW at low temperatures is absent or weak, however, this doesn't automatically imply that a CDW order is absent above $T_c$ in Bi2212. We will discuss the pseudogap above $T_c$ further.

The excitation gap with the maximum at ($\pi/2$, $\pi/2$) is a spin-excitation gap which occurs inside the SDW gap. However, the spin-excitation gap has a very strange behavior. It behaves like a SC gap but there is no Josephson current in tunneling spectra corresponding to this gap. There is only one explanation of this remarkable fact. The spin-excitation gap is a SC gap which corresponds to the bound state of fermionic spinons. Since there is separation of spin and charge in a locally striped structure [3] as in the one-dimensional (1D) electron gas [5], the excitation gap is a pure spin gap. The charge is carried by bosonic holons. There is no gap in the charge response. Hereafter, we will call the excitation gap as the spinon-SC gap.

We discuss now the dependence of the magnitudes of the spinon-SC and SDW gaps in Bi2212 on the hole concentration, $p$. It is noteworthy that the magnitude of the spinon-SC gap is approximately one fourth of the magnitude of the SDW gap which is a normal-state gap (see, for example, Fig. 2 of Ref. 26). Figure 9 shows the phase diagram for the spinon-SC gap and SDW gap. The hole concentration has been obtained from the empirical relation $T_c/T_{c,\,max} = 1 - 82.6(p - 0.16)^2$ which is satisfied for a number of HTSC and we use $T_{c,\,max} = 95$ K. Measurements have been performed on underdoped Bi2212 single crystals with $T_c = 21$ K (one sample), 51 K, and 76 K and on overdoped samples which were described above. The errors of the measured data in Fig. 9 for the spinon-SC gap at 14 K are small ~ $2(k_B T) = 2.4$ meV [18]. For the spinon-SC gap, we find a good agreement with tunneling data also obtained by a B-J technique [26] and shown in Fig. 9.

## 6 The D-wave superconducting gap

From the Section 4, we know that the tunneling gap in Bi2212 below $T_c$ is the combination of a SC gap having the Josephson current, a spinon-SC gap and SDW gap. One can see in Fig. 5 (keeping in mind Fig. 6) that the Josephson current is absent in $(\pi, \pi)$ direction on the Fermi surface. Consequently, the SC gap is absent at $(\pi/2, \pi/2)$. The Josephson current is maximum at $(\pi, 0)$. The two gaps, the SC gap and spinon-SC gap can be distinctly seen in the spectrum C shown in Fig. 5. The spinon-SC gap is always larger than the SC gap. The maximum of the SC gap is located at $(\pi, 0)$. Femtosecond time-resolved spectroscopy [31] has revealed an evidence for the existence of two components of the HTSC: band-like and polaronic-like carriers.

Figure 10 shows $I(V)$ and $dI/dV(V)$ characteristics measured at 15 K on a Ni-doped Bi2212 single crystal with $T_c = 75$ K. The temperature dependence of the $dI/dV(V)$ is presented in Fig. 11. The $I(V)$ characteristic in Fig. 10 is *remarkable*, it is almost flat outside of the gap. The shapes of the $I(V)$ curve and the conductance curve (with the exception of the Josephson current) look very similar to the shape of a polaron (or bipolaron) [30] and similar to the curves A - B shown in Figs. 7 and 8. The conductance curve almost doesn't have the background. A very high value of the Josephson current $I_J = 0.11$ mA and the temperature dependence of the measured DOS shown in Fig. 11 indicate the SC nature of the spectra.

The explanation is as follows. Most likely, the distribution of Ni in this sample was not uniform. We tested the DOS in the vicinity of a small cluster of $Ni^{2+}(S = 1)$. Ni is most likely a magnetic impurity in Bi2212 [32, 33]. Locally, spins $S = 1$ destroy partially the AF order. Consequently, it destroys partially the SDW gap. That is why the background of the conductance curve is so low. According to the Anderson's theorem a magnetic impurity will destroy locally a s-wave SC. The spins $S = 1$ of Ni will allow to propagate in it's vicinity only excitations with $S = 1$. Thus, the spectrum in Fig. 10 can only be explained in two ways: it is a triplet (p-wave) SC state, or it is a d-wave SC state mediated by spin-waves. Since we know that the SC in cuprates has the predominant d-wave character, we find that in cuprates there exists a d-wave SC mediated by spin-waves with $\Delta_{d,max} < \Delta_{sp,max}$, where $\Delta_{d,max}$ and $\Delta_{sp,max}$ are maximum magnitudes of the d-wave and spinon-SC gaps, respectively. The maximum of the d-wave gap is located at $(\pi, 0)$ on the Fermi surface.

In the Introduction, we mentioned recent inelastic neutron scattering (INS) experiments. Here we will show that the resonance peak observed in the SC state of YBCO [6,7,4] corresponds to spin-waves from the d-wave SC. We know that in order to break a Copper pair we need an energy of $2\Delta$. Let's assume that the frequency of the resonance peak detected by INS is equal to $2\Delta$. Figure 12 shows INS data for YBCO [6,7] and our tunneling data for Bi2212. One can see that there is a good agreement between two sets of data obtained from different types of cuprates. Thus, the INS detects spin-waves which are an essence of a SC of the magnetic origin. It is obvious why in INS measurements the resonance peak shifts to higher frequencies with decreasing temperature.

In Fig. 8 we observe a pure d-wave gap of magnetic origin with the value of $2\Delta_d(p = 0.11) = 34$ meV. However, the d-wave gap in the tunneling spectra shown in Fig. 5 is detected in the presence of the spinon-SC and SDW gaps. As shown above (see Fig. 8), when the SDW gap is subtracted from the measured spectrum the value of the tunneling gap becomes a little bit smaller. So, it is reasonable to accept a value of $2\Delta_d(p = 0.19) = 44$ meV (see Figs. 1 and 5) for the pure d-wave component in overdoped Bi2212 ($p = 0.19$).

This is for the first time a direct evidence of a $d_{x^2-y^2}$ SC mediated by spin-waves is presented in the literature. At the same time, it is one more piece of evidence that the large non-SC gap below $T_c$ has the predominant magnetic origin, *i. e.* it is a SDW gap. It seems that a CDW at low temperatures is absent or weak.

# 7 The sub-gap structure

Figure 13 depicts all spectra shown in Fig. 5 together (overlapped). One can note that the spectra in Fig. 13 (spectra D and E shown in Fig. 5) have a sub-gap structure. Figure 14 shows the temperature dependence of the spectrum F in Fig. 5. It is evident that the sub-gap of the spectra D and E in Fig. 5 can still be seen in Fig. 14 at 29 K. One can see in Fig. 5 that the sub-gap develops with the appearance of the d-wave magnetic-SC gap. A weak sub-gap structure is also present in purely magnetic spectra shown in Fig. 11, which looks similar as in Fig. 14. Thus, it is logical to associate the sub-gap with the magnetic SC. Theoretically, a d-wave magnetic SC may coexist with a g-wave magnetic SC [14,15]. The magnitude of the sub-gap is approximately a half of the d-wave gap and it depends very weakly on the hole concentration while the d-wave component has a perceptible dependence. It is in a good agreement with the theory of a d-g-wave magnetic SC [34]. Further, the sub-gap will be considered as the g-wave magnetic gap. In fact, the exact symmetry of the sub-gap is not very important. It is a least what we have to worry about.

## 8 Relations among different OPs

In Bi2212, there are 4 different gaps at low temperatures, namely: (i) a SDW gap, $\Delta_{SDW}$; (ii) a spinon-SC gap, $\Delta_{sp}$; (iii) a d-wave SC gap of the magnetic origin, $\Delta_d$, and (iv) a small g-wave SC gap of the magnetic origin, $\Delta_g$. Maximum magnitudes of the gaps relate to each other in a simple way: $\Delta_g < \Delta_d < \Delta_{sp} < \Delta_{SDW}$. The angle between lobes of the d-wave OP and the spinon-SC OP is equal to 45°. In Fig. 13, one can compare visually the intensities of the four gaps in overdoped Bi2212 ($p = 0.19$). The d-wave component (spectrum F in Fig. 5) is the strongest. It is 1.5 times more intense than the spinon-SC component (spectrum A in Fig. 5) which can have a s-wave or d-wave symmetry [5]. Indeed, the SC in the cuprates has the predominant d-wave character.

Now we discuss temperature dependencies of the SC gaps. Figure 15 shows temperature dependencies of tunneling spectra for Bi2212. The curve A displays a temperature dependence of the spectra shown in Fig. 11, which have a purely magnetic origin. The curve B in Fig. 15 corresponds to the spectra shown in Fig. 14. The curve C is a *typical* temperature dependence for a maximum tunneling gap or for a gap which is close to the maximum value [24, 26], *i. e.* this is a temperature dependence of the spinon-SC gap with the presence of the SDW gap. Both, B and C temperature dependencies shown in Fig 15 develop above $T_c$ into the pseudogap [22, 26]. Consequently, since the magnetic SC (curve A) disappears with increase of temperature, it is clear that the spinon-SC component evolves into a pseudogap gap above $T_c$. The chain of events during the cooling down is next: the development of a SDW gap - the appearance of a spinon SC - the appearance of a d-wave magnetic SC. The temperature dependence of the g-wave gap is similar to the curve A in Fig. 15 (see, for example, Fig. 14).

From temperature dependencies C and B shown in Fig. 15, we conclude that there is one SC gap at ($\pi/2$, $\pi/2$) on the Fermi surface. It is logic since the magnitude of the d-wave SC gap is equal to zero at ($\pi/2$, $\pi/2$). The second conclusion is that there are two SC gaps at ($\pi$, 0). In two-band SC, the temperature dependence of one of the two gaps lies below the BCS temperature dependence [18]. This implies that the spinon-SC gap is non-zero at ($\pi$, 0) on the Fermi surface. This points out that the spinon-SC gap does not have a pure d-wave symmetry. It has either a pure s-wave or mixed (s+d) symmetry. This is an important conclusion.

Finally, we present the phase diagram for the SC gaps. Figure 16 shows maximum values of the SC gaps measured at 14 - 15 K as a function of the hole concentration. We have reanalyzed the data for the spinon-SC gap presented in Fig. 9. The magnitude of the SC gap for the $T_c$ = 21 K sample can not be associated either with the spinon or magnetic SC. Most likely, the oxygen distribution in the $T_c$ = 21 K sample was not homogeneous. So, in Fig. 16 we do not use the data for the $T_c$ = 21 K sample. The errors of measured data along the vertical axes in Fig. 16 correspond to the size of dots in Fig. 16. The dashed line for the spinon-SC gap is a straight line. It seems that the spinon SC exists on stripes below $p$ = 0.05 and, probably, above $p$ = 0.27. However, the magnetic SC disappears at these points. Thus, the points (0.05, 0) and (0.27, 0) for the d-wave and g-wave magnetic gaps in Fig. 16 are assumed from this fact that $T_c$ = 0. Most likely, the magnitudes of the d-wave and g-wave gaps have maxima at optimal doping. The dependence of the magnitude of the SDW gap on hole concentration is presented in Fig. 9.

## 9 The MS model

In this Section, we will use the purely experimental data discussed above to construct a model which fits the data. We will take into account also some data obtained by other techniques.

It has been shown above that the $\Delta_{sp}$ is always larger than the $\Delta_d$. Since, $\xi \sim 1/\Delta$, consequently, $\xi_d > \xi_{sp}$, always. *This is the clue.* The whole HTSC scenario can be compressed into one phrase: the spinon SC is a SC on stripes, the magnetic SC is a SC between stripes. We will call the model for the HTSC as a magnetic-spinon (MS) superconductivity (or a magnetic-stripes superconductivity). There are 3 characteristic temperatures in the MS model: $T_c < T_{pair} < T_s^*$, where $T_{pair} = T_2^*$ and $T_s^* = T_1^*$ (see Section 3). The $T_s^*$ and $T_{pair}$ correspond to the formation of charge stripes and to the pairing of spinons, respectively. In fact, our scenario above $T_c$ is very similar to the scenario above $T_c$ described by Emery, Kivelson and Zachar [5].

Let's consider three different temperature regions.

1) $T_{pair} < T \quad T_s^*$. Holes doped into the $CuO_2$ planes form charged stripes which separate AF domains [3,36]. The $T_s^*$ is a temperature at which charge stripes are formed, which depends on hole concentration. Above $T_s^*$, holes are uniformly distributed into the $CuO_2$ planes.

2) $T_c < T \quad T_{pair}$. The spin-spin interactions of neighboring AF domains are not cut off completely by the charged stripes [37]. So, a large spin-pseudogap arises naturally between neighboring AF zone boundaries. In a locally striped structure, there is a separation of spin and charge. At $T_{pair}$, spinons on an individual stripe acquire a spin-gap due to the spin-pseudogap. The spinon anisotropic spin-gap is defined by the anisotropic spin-pseudogap. As we concluded in the previous Section, the spinon-SC gap has a s-wave or (s+d) mixed symmetry. The maximum of the spinon-SC gap is located along lines 45° away from the Cu-O bond direction where spin-spin interactions in the $CuO_2$ planes are maximum. The spinon coherent length is too short in comparison with the distance between stripes in order to establish a coherent SC state in a whole volume of a sample at high temperatures. The $T_{pair}$ depends on hole concentration and scales with the $T_s^*$. The spinon SC and stripe order coexist at $T_c < T \quad T_{pair}$.

At any temperature, quasiparticles and separate holes (electrons) tunnel incoherently from one stripe to another. Tunneling through a medium with the AF order, they excite spin-waves (magnons) [2]. Above $T_c$, the spin-waves are damped. The tunneling hole (electron) can be considered as a magnetic polaron [2]. Due to strong magnon-electron and magnon-magnon interactions [2], two holes (electrons) can communicate with each other by means of spin-waves. An alternative explanation of magnetic polarons can be that on charged stripes there exists an optimal charge density [3] which occurs at $p = 0.05$. The stripes try to sustain this optimal charge density. The charge exceeding the optimal density is expelled into AF domains.

3) T $\quad T_c$. At $T_c$, the magnetic SC occurs. A d-wave SC mediated by spin-waves is well described by spin fluctuation theories [8-16]. The spinon SC develops into the coherent SC state due to the magnetic SC. Thus, the coherent SC state in a whole volume of a sample for the spinon SC is established via the magnetic SC and not by the Josephson coupling between the stripes like in NCCO which we will discuss further. The coherent length of the magnetic SC is defined by the spacing between stripes and not by the range of the pair wave function in BCS theory. The d-wave magnetic SC appears due to the existence of walls made out of electron pairs. It should be understood that the spinon SC on stripes could develop into a coherent SC state without the magnetic SC but it would occurs at much low temperatures like in NCCO.

In NCCO, the OP has most likely an anisotropic s-wave symmetry [35]. This implies the absence of the magnetic SC. The reason for this is most likely the Nd spins which couple to the Cu subsystem [38] and suppress spin-waves. The spinon SC in the absence of the "partner" develops alone into the coherent state due to the Josephson coupling between the stripes. The critical temperature is lower than in $La_{2-x}Sr_xCuO_4$ where the $T_c$ is determined by mutual efforts of the spinon SC and the magnetic SC. The coherence length of Copper pairs in NCCO (~70 - 80 Å) has to be larger than the distance between stripes. The average distance between stripes is easy to estimate. It is equal to the *ab*-coherence length (in YBCO, Bi2212, *etc.*), consequently, d ~ 10 - 25 Å. It is possible also that the saddle-point in the DOS at ($\pi$, 0) on the Fermi surface, which produces a Van Hove singularity [39], plays an important role in the appearance of the d-wave magnetic SC. In NCCO, there is a large energy separation between the saddle-point and the Fermi level.

There is an excellent piece of evidence in favor of the MS model. Rener *et. al.* [40] have observed by STM a normal-state pseudogap in the vortex cores of Bi2212 at low temperatures, *i. e.* the spinon SC. It is logic that $B_{c1,\text{s-wave}} > B_{c1,\text{d-wave}}$ since $\Delta_s > \Delta_d$. Inside the vortex cores the magnetic SC was suppressed while the spinon SC was still "alive". In the vortex cores of YBCO at low temperatures they have observed a large gap corresponding to the spinon SC on stripes like in Bi2212 and a smaller gap. It seems that this smaller gap corresponds to the SC on chains (!) The chain SC occurs also due to pairing of spinons.

Finally, it must be underlined that the high-$T_c$ superconductivity, as a phenomenon, consists of a chain of events: the formation of charged stripes - the development of a spin-pseudogap due to AF correlations - the appearance of a spinon SC - the appearance of a d-wave magnetic SC. The charged stripes are the intermediate steps leading to the high $T_c$. The main difference between NCCO and other cuprates is that the coherent SC state is established differently: due to the Josephson coupling between stripes and due to the magnetic SC, respectively.

## 10 Discussion

In the context of the MS model, let's consider some important issues regarding the HTSC.

*The pseudogap above $T_c$.* We have showed above that the normal-state gap below $T_c$ is a predominant SDW gap. Above $T_c$, the pseudogap is a combination of the SDW gap and most likely a CDW gap. Below $T_{\text{pair}}$, a spin-gap due to pairing of spinons, which develops inside the SDW gap, is the addition to the pseudogap. The spinon-SC gap can be considered *in some sense* as a precursor pairing.

*The $T_s^*$ value.* In overdoped Bi2212, it corresponds to the value of $T_s^* = 280 - 290$ K (see Section 3). For underdoped Bi2212, the $T_s^*$ is higher, $T_s^* > 300$ K [27,22].

*The $T_{\text{pair}}$ value.* In overdoped Bi2212, it is of the order of 130 K and $T_{\text{pair}} \sim 180 - 190$ K in underdoped Bi2212 [20].

*The 1/8 problem.* This is connected with the disappearance of one of the two SCs. It is most likely that it is the spinon SC which disappears. If it would be the magnetic SC, the spinon SC would develop into a coherent SC state at lower temperatures like in NCCO. The reason for the absence of the spinon SC can be only the absence of the AF correlations at $p = 0.125$. Most likely, this corresponds to a glassy state [37].

*The $2\Delta(0)/k_B T_c$ value.* There is no sense to calculate it for the spinon SC gap. For the magnetic d-wave SC gap, the $2\Delta(14\text{ K})/k_B T_c$ value in Bi2212 for doping levels, $p = 0.11$ ($T_c = 75$ K) and $p = 0.19$ ($T_c = 89$ K) is equal to 5.3 and 5.7, respectively. For YBCO, if we use INS data at low temperature, the $2\Delta/k_BT_c$ value for single crystals with $T_c$ = 52 K and $T_c$ = 67 K [7] is equal to 5.6 and 5.7, respectively.

So, it seems that the $2\Delta/k_BT_c$ value in cuprates is of the order of 5.7. If we assume that the magnitude of the d-wave magnetic gap shown in Fig. 16 has a maximum at optimal doping and we use $T_{c,max}$ = 95 K and $2\Delta/k_BT_c$ = 5.7 in Bi2212, then we obtain $\Delta_{d,max}$ = 23.3 meV.

The magnitude of the d-wave magnetic SC gap in cuprates (with the exception of NCCO) can be expressed as $\Delta/\Delta_{max} = 1 - 82.6(p - 0.16)^2$. In Bi2212 and YBCO, the $\Delta_{max}$ is equal to 23.3 meV and 23 meV, respectively. In LSCO, if we use $T_{c,max}$ = 38 K, the $\Delta_{max}$ is equal to 9.3 meV.

*Why does everything depend on p?* Because the value of the spin-pseudogap is determined by the hole concentration. The value of the spinon SC gap is defined in a unique fashion by the value of the spin-pseudogap. The value of the magnetic d-wave gap depends on the distance between stripes and consequently on the hole concentration. Thus, the hole concentration is an unique parameter for bulk HTSC characteristics. We expect that the phase diagram shown in Fig. 16 will be similar for other cuprates with different scales on the vertical axis. However, the phase diagram will be slightly different for YBCO (because of chains) and, obviously, different for NCCO.

*Why is $T_c$ so high?* Due to 1D physics and the high value of $J = 0.15$ eV.

*What is the origin of the HTSC in one word?* The HTSC as a phenomenon is a multi-step process. Let's consider the origin of each intermediate step. The phase separation occurs due to a metal-insulator transition. The development of the spin-pseudogap has a magnetic origin. The spinon SC is a consequence of the two previous steps. The origin of the magnetic SC is obvious. So, taken all into account the origin of the HTSC is more justified to be called magnetic than any other name for it.

## 11 Andreson's theorem for cuprates

By analogy with Anderson's theorem for classical superconductors, we can formulate a similar theorem for cuprates: *any impurity doped into $CuO_2$ planes will destroy the superconductivity.*

From Section 9, we know that the HTSC consists of a chain of events: striped phase - spin-pseudogap - singlet spinon SC - d-wave magnetic SC. Any defects at any initial stage of this chain will affect the final result. Any impurity will destroy, at least, one element of this chain. For example, a magnetic impurity will locally damage, partially, the spin-pseudogap and completely destroy the s-wave spinon SC. A non-magnetic impurity will locally destroy the spin-pseudogap and the magnetic SC. It must be underlined that the spin-pseudogap is the most important element in this chain. Both, a magnetic and non-magnetic impurities damage it. The theorem is valid for NCCO too. Even, if a doped impurity will be in a $(S = 1/2)^{2+}$ state, it will affect the SC due to the difference in ionic radius with $Cu^{2+}$, for example, *etc*.

## 12 Conclusions

The evidence of a $d_{x^2-y^2}$ SC mediated by spin-waves is presented for the first time in the literature. The Magnetic-Spinon model for the high-$T_c$ superconductivity also for the first time is presented in the literature. Our data are in a good agreement with other experimental data obtained by other techniques on different types of cuprates.

Finally, I summarize the content of present work. The complete scenario of high-$T_c$ superconductivity based on experimental data from electron-tunneling spectroscopy on underdoped, overdoped and Ni-doped $Bi_2Sr_2CaCu_2O_{8+x}$ single crystals in a temperature range between 14 K and 290 K using a break-junction technique is presented. Below $T_c$, we observe four different gaps, namely, (i) a SDW gap due to antiferromagnetic correlations; (ii) a superconducting gap due to spinon pairing; (iii) a d-wave superconducting gap, and (iv) a small superconducting gap most likely having g-wave symmetry. We show that the d-wave superconductivity is mediated by spin-waves and the magnitude of the superconducting gap due to pairing of spinons is larger than the magnitude of the d-wave gap, however, the d-wave gap is more intense. The superconducting gap due to pairing of spinons most likely has a s-wave symmetry. The maximum magnitudes of the SDW, the spinon and d-wave gaps are located at ($\pi/2$, $\pi/2$), ($\pi/2$, $\pi/2$) and ($\pi$, 0) on the Fermi surface, respectively. The d-g-wave superconductivity mediated by spin-waves can be considered as a pairing of magnetic polarons. Thus, there are two different types of superconductivity in cuprates with the exception of $Nd_{2-x}Ce_xCuO_4$ (NCCO), where there is only the spinon superconductivity. The main difference between NCCO and other cuprates is that the spinon coherent superconducting state is established differently: due to the Josephson coupling between stripes and due to the magnetic superconductivity, respectively. The presented MS model naturally explains other experimental data. We formulate a theorem for cuprates by analogy with the Anderson's theorem for classical superconductors.

I thank R. Deltour, J. Zaanen, O. P. Sushkov and D. van der Marel for discussions and H. Hancotte for help with experiment. This work is supported by PAI 4/10.

**Figure captions**

FIG. 1. Tunneling spectra for a SIS junction measured as a function of temperature on an overdoped Bi2212. The conductance scale corresponds to the 290 K spectrum, the other spectra are offset vertically for clarity. The curves have been normalized at -150 mV (or nearest point).

FIG. 2. Tunneling spectra measured as a function of temperature on an overdoped Bi2212. The conductance scale corresponds to the 290 K spectrum, the other spectra are offset vertically for clarity. The curves have been normalized at -200 mV (or nearest point).

FIG. 3. The odd conductance, - $G$o($V$) for the spectra shown in Fig. 2 as a function of temperature. The conductance scale corresponds to the 290 K spectrum, the other spectra are offset vertically for clarity.

FIG. 4. Temperature dependence of the odd conductance - $G$o($V$) shown in Fig. 3. The dashed line is a guide to the eye.

FIG. 5. Tunneling spectra measured at 14 K on an overdoped Bi2212 single crystal with $T_c$ = 89.5 K. The conductance scale corresponds to spectrum A; the other spectra are shifted vertically for clarity by 1.2 units from each other. The spectra are normalized at -150 meV.

FIG. 6. Tunneling gap at low temperature vs. angle on the Fermi surface: dots (average measured points) and solid line (an assumption of a fourfold symmetry). The data are taken from Ref. 29.

FIG. 7. Tunneling spectra A and B measured at 14 K on an underdoped Bi2212 single crystal with $T_c$ = 51 K. The spectra are normalized at - 400 meV. The spectrum A - B is the difference between spectra A and B. The conductance scale corresponds to spectrum B; the spectra A and A - B are shifted vertically for clarity (zero conductance is indicated for each spectrum by the horizontal line at zero bias).

FIG. 8. Tunneling spectrum A measured at 14 K on an overdoped Bi2212 single crystal with $T_c$ = 88 K. The spectrum is normalized at -150 meV. The spectrum B is created from the spectrum A (see text) and corresponds to a normal-state gap. The spectrum A - B is the difference between spectra A and B. The conductance scale corresponds to spectrum A; the spectra B and A - B are shifted vertically for clarity by -0.05 and +1.2 units, respectively. The $R_N$ of the junction is equal to 300 kΩ.

FIG. 9. Maximum gap values at 14 K vs. hole concentration in Bi2212 single crystals: diamonds and squares (SDW gap at ($\pi/2$, $\pi/2$)); dots (spinon-SC gap at ($\pi/2$, $\pi/2$)), and triangles (maximum tunneling gap, Ref. 26). The points (0.16, 150) and (0.23, 83) for the SDW gap (squares) are taken from Refs. 25 and 27, respectively. The dashed lines are guides to the eye.

FIG. 10. *I(V)* and *dI/dV(V)* characteristics measured at 15 K on a Ni-doped Bi2212 single crystal with $T_c$ = 75 K. The conductance curve is normalized at -200 meV.

FIG. 11. Normalized *dI/dV(V)* characteristics vs. temperature for the Ni-doped Bi2212 single crystal (see Fig. 10). The conductance scale corresponds to the 70.3 K spectrum, the other spectra are offset vertically for clarity. The conductance curves are normalized at -150 meV (or nearest point).

FIG. 12. The magnitude of the d-wave magnetic SC gap in Bi2212 (triangles) and neutron scattering data on YBCO (dots) at low temperatures [6,7] vs. hole concentration.

FIG. 13. The tunneling spectra from Fig. 5.

FIG. 14. Tunneling spectrum F shown in Fig. 5 measured as a function of temperature. The conductance scale corresponds to the 104 K spectrum, the other spectra are offset vertically for clarity. The curves have been normalized at -100 mV (or nearest point).

FIG. 15. Measured temperature dependencies of the quasiparticle DOS in Bi2212 single crystals. The dependence A corresponds to the pure d-wave magnetic-SC gap (without background) shown in Fig. 11. The dependence B corresponds to the spectra shown in Fig. 14. The dependence C is a typical temperature dependence for a maximum SC gap [24, 26].

FIG. 16. Maximum SC gap values at 14 - 15 K vs. hole concentration in Bi2212 single crystals: dots and diamond (spinon-SC gap at ($\pi/2$, $\pi/2$)); triangles (d-wave magnetic-SC gap at (0, $\pi$)), and squares (g-wave magnetic-SC gap at (0, $\pi$)). The point (0.23, 21) for the spinon-SC gap (diamond) is taken from Ref. 27. The dashed lines are guides to the eye.

**(NOTE: the figures are in .ps format with poor quality for graphs)**

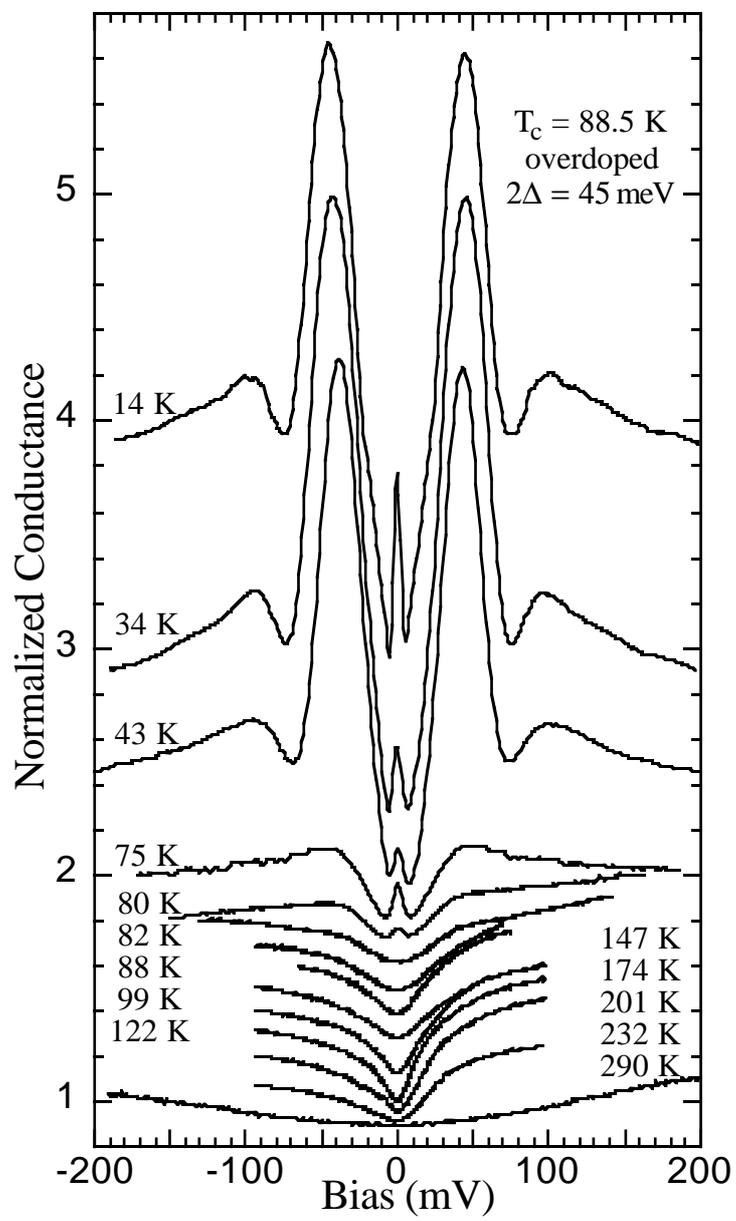

FIG.1

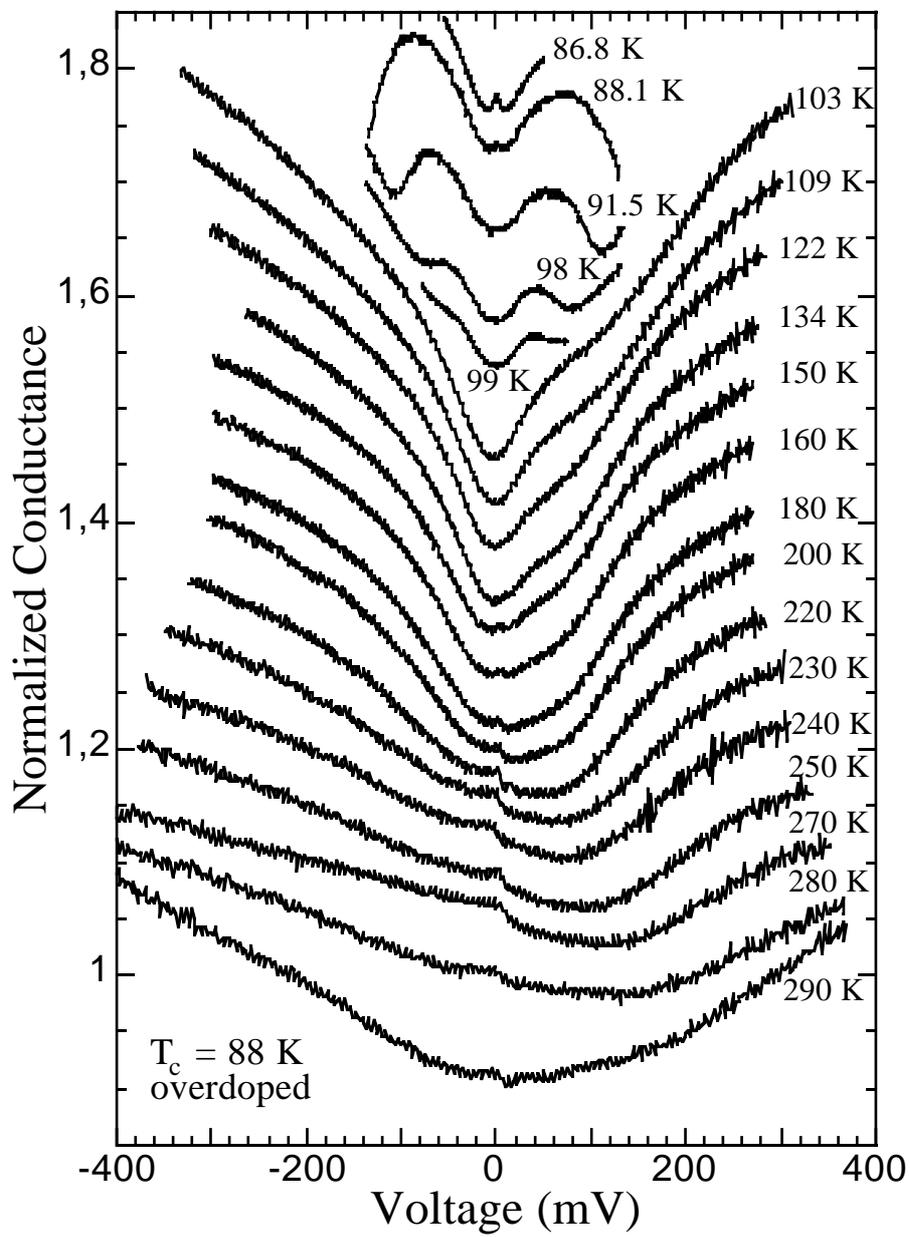

FIG. 2

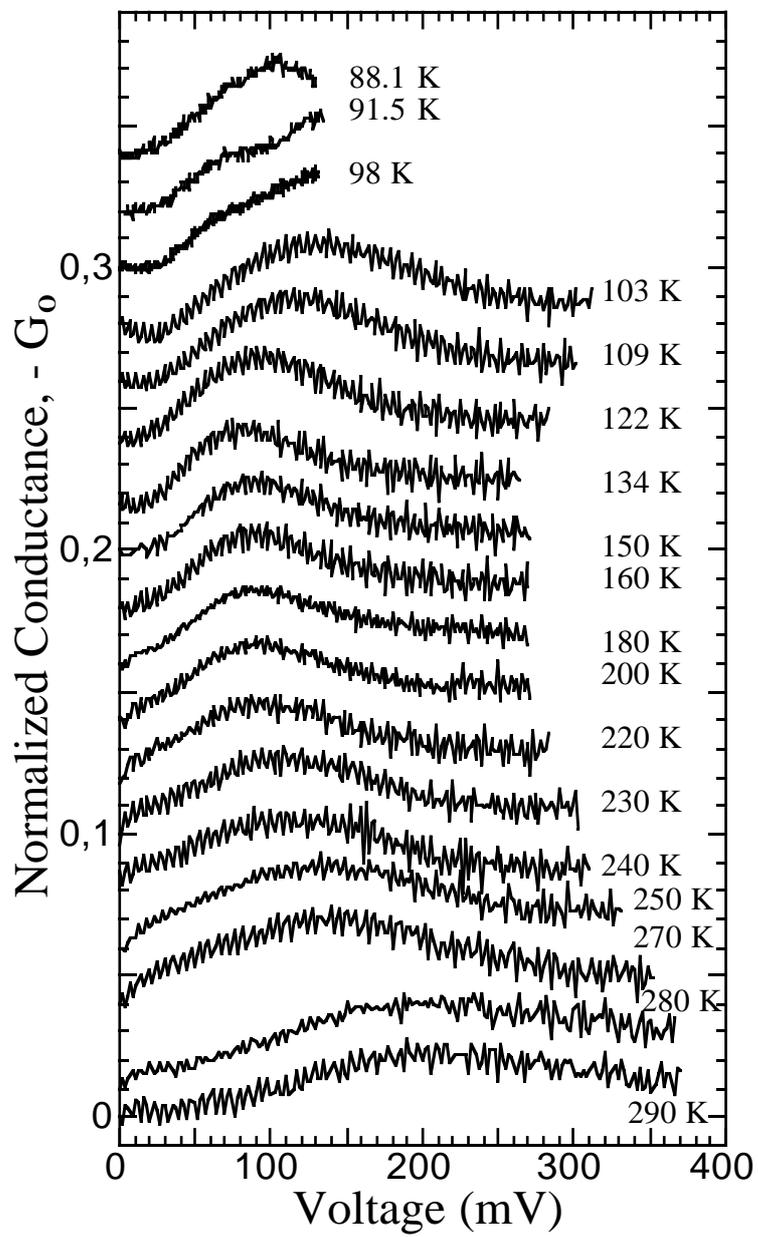

FIG. 3

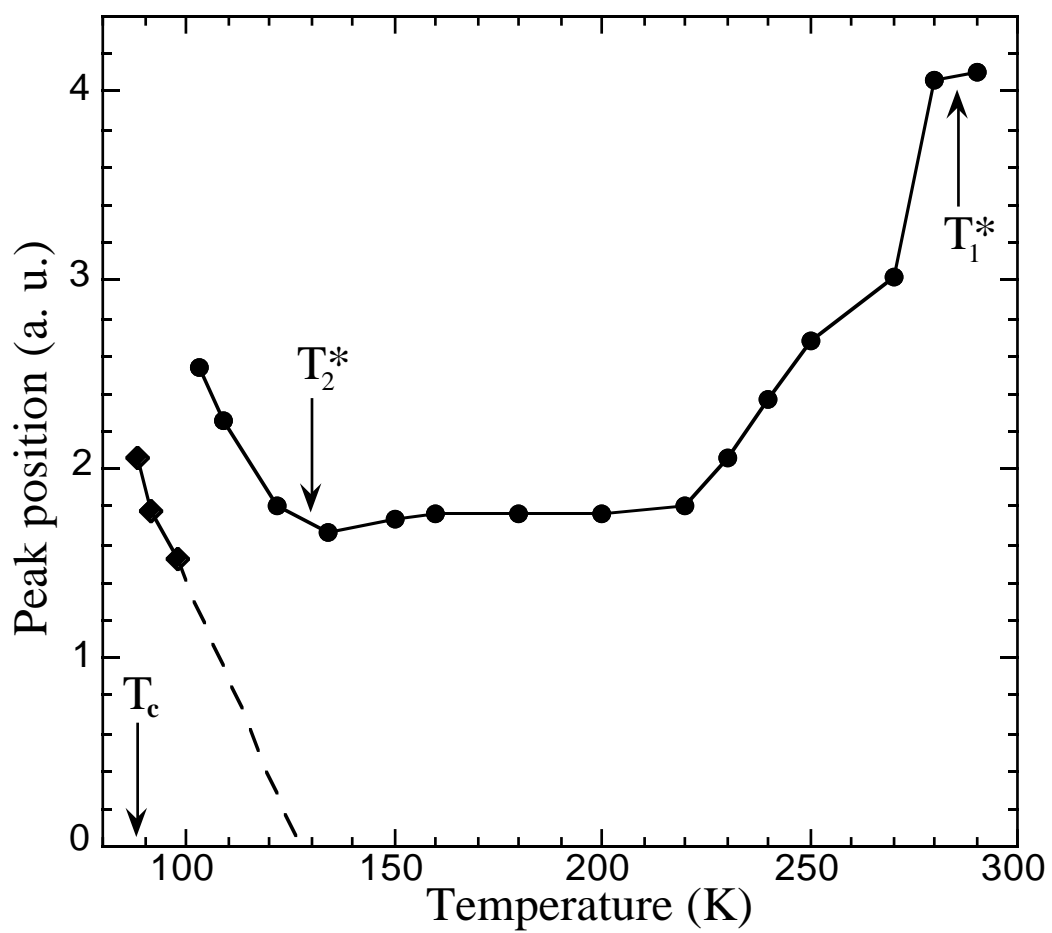

FIG. 4

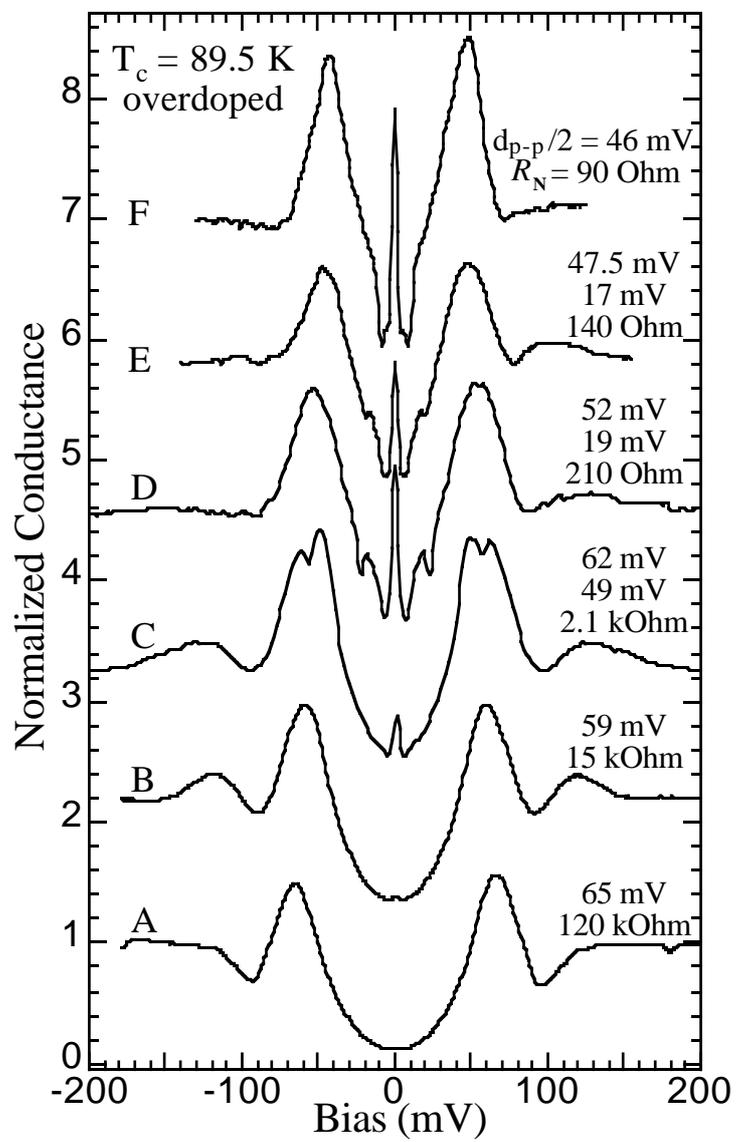

FIG. 5

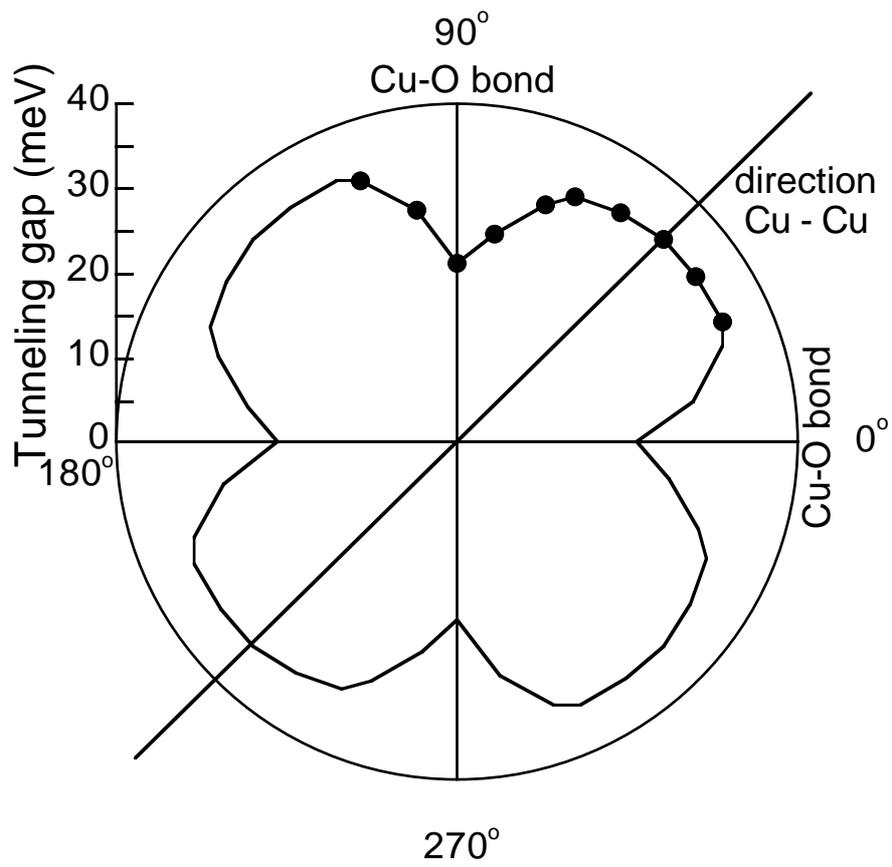

FIG. 6

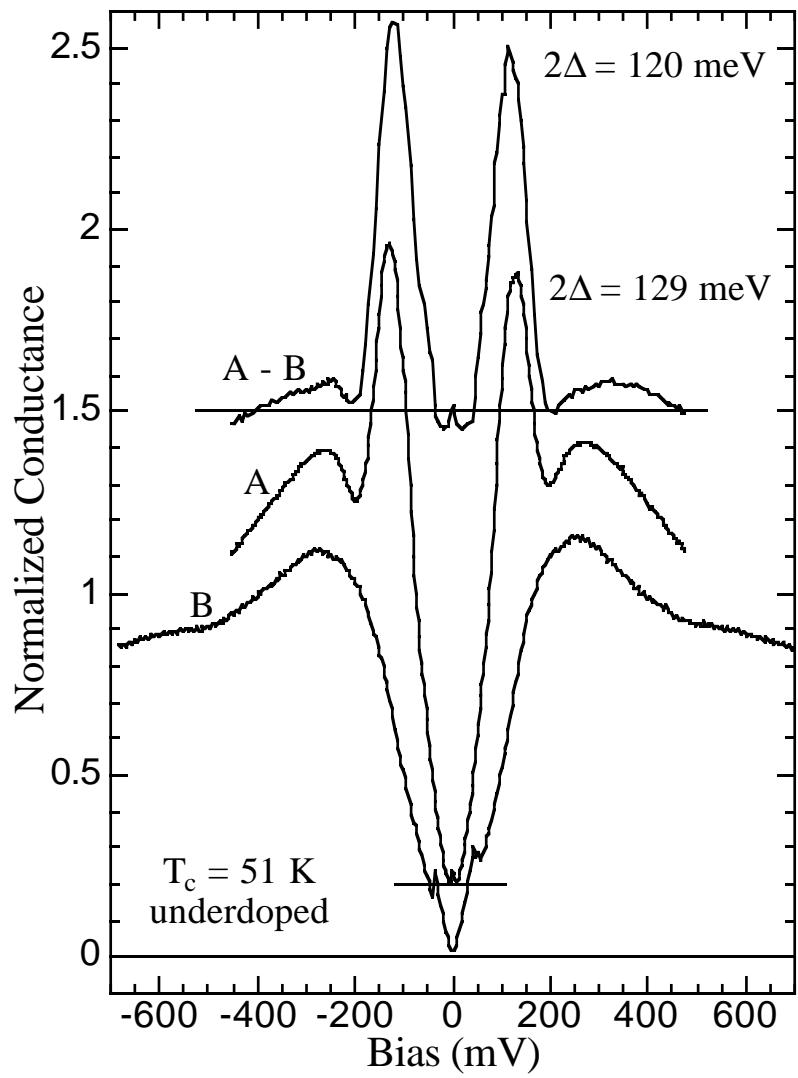

FIG. 7

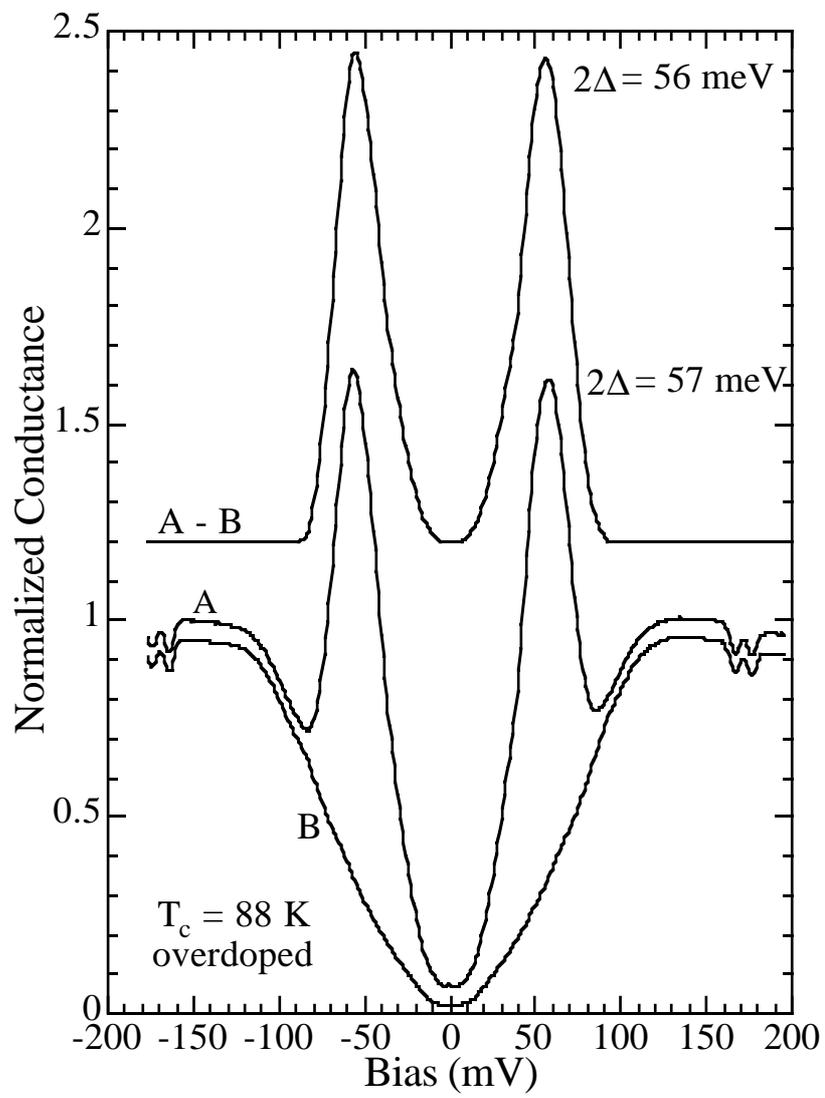

FIG. 8

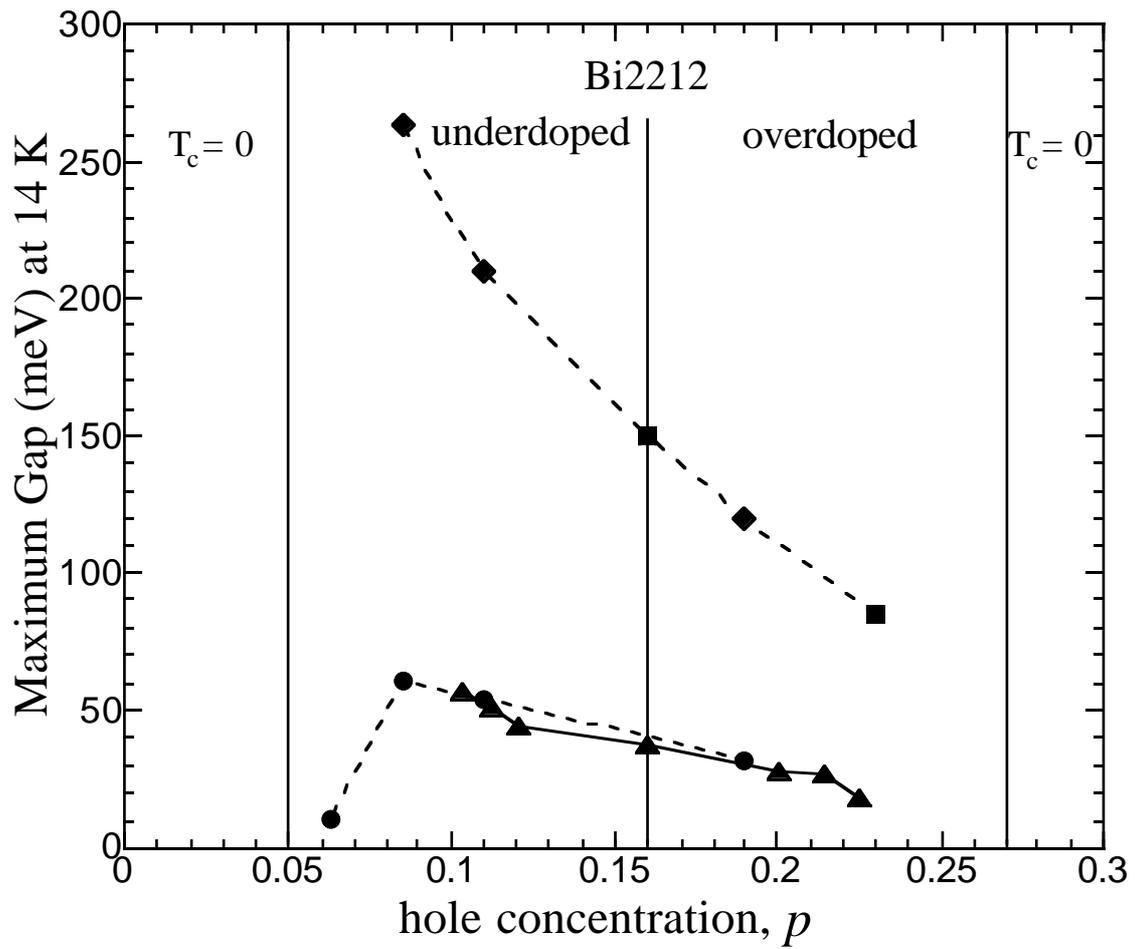

FIG. 9

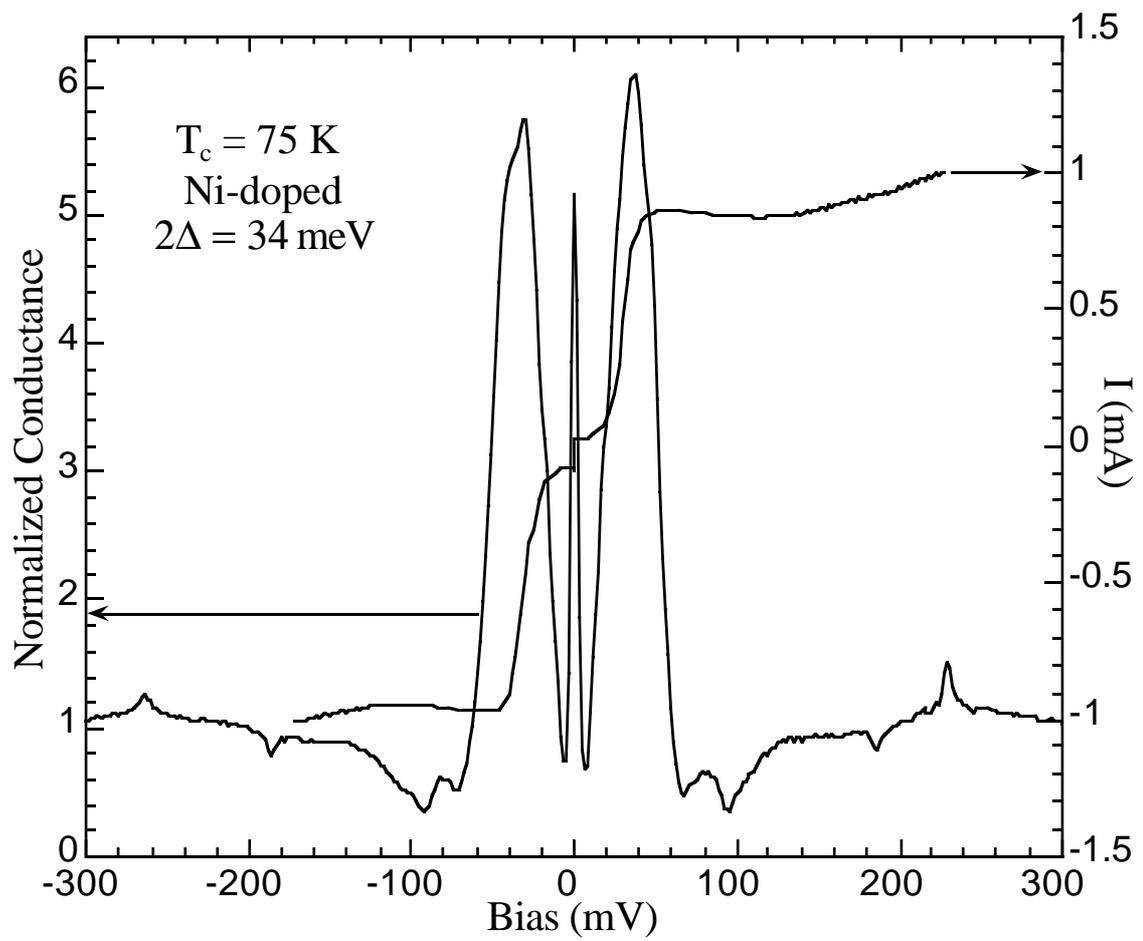

FIG. 10

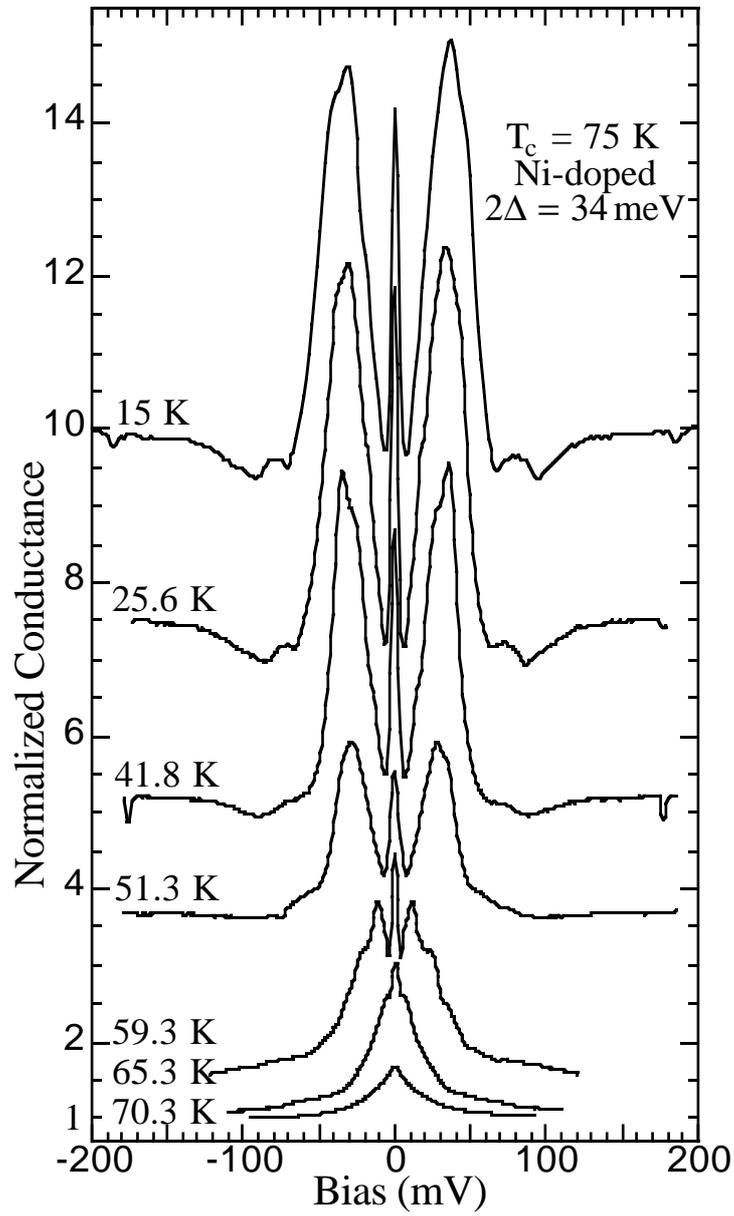

FIG. 11

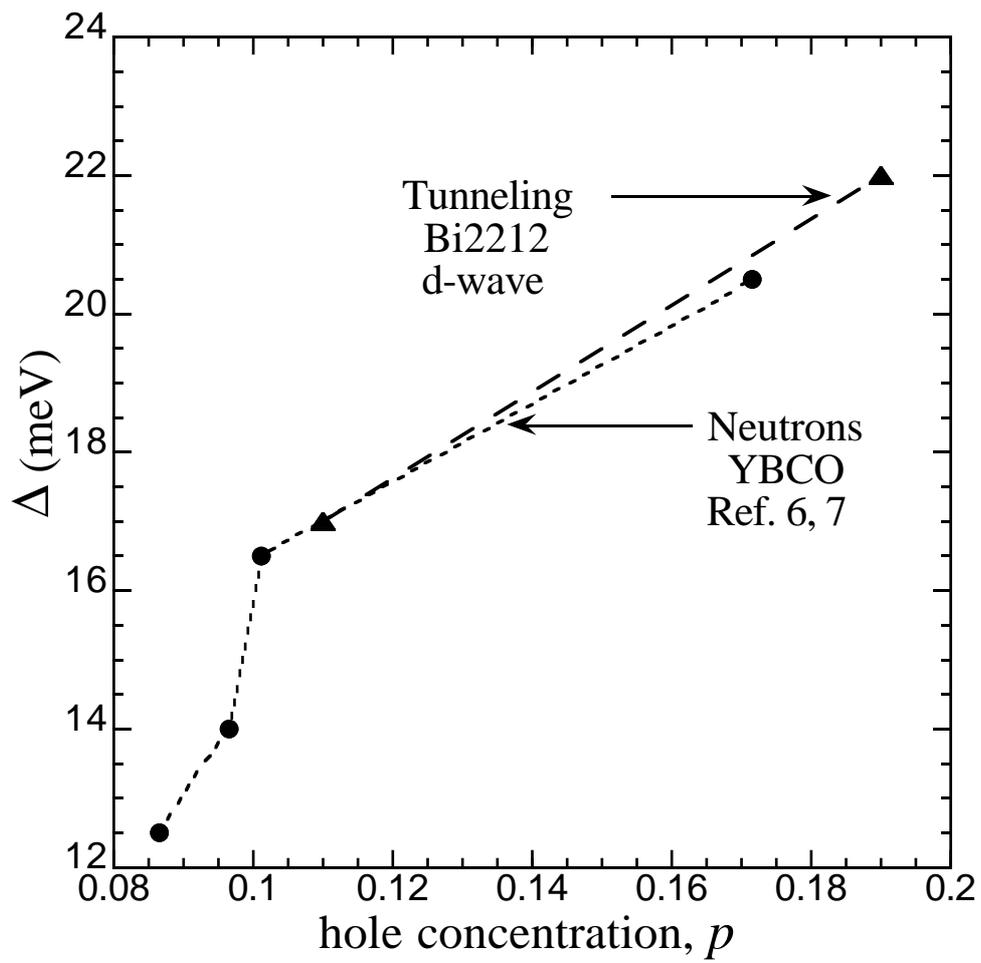

FIG. 12

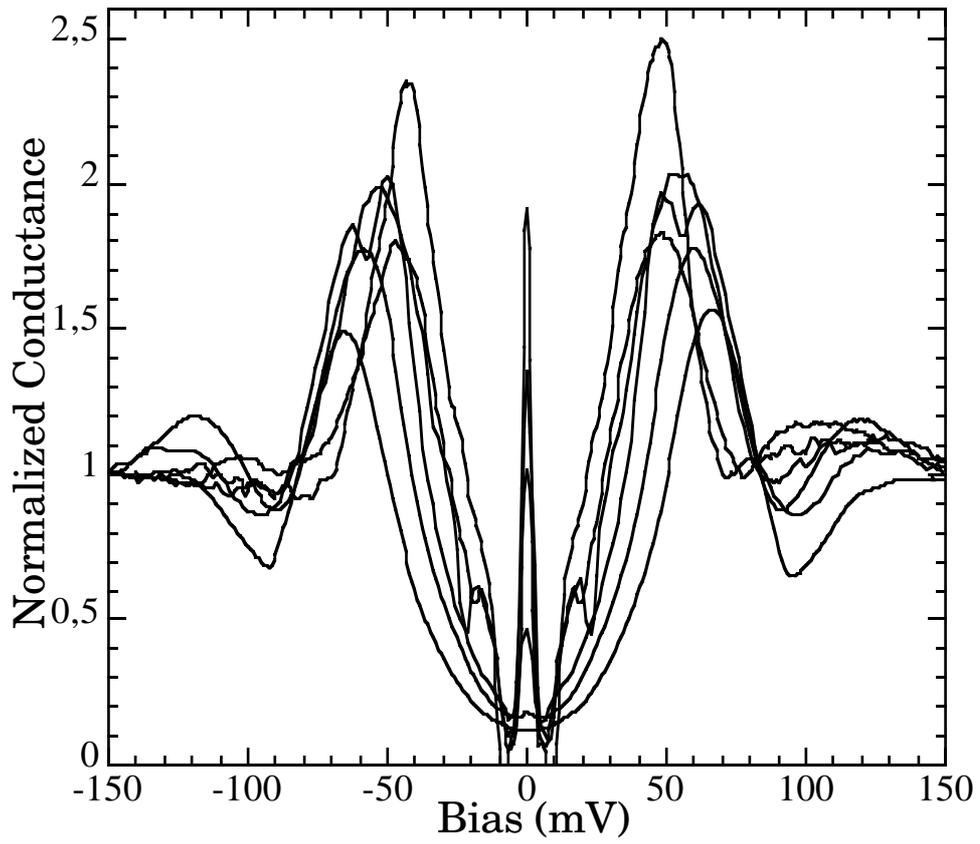

FIG. 13

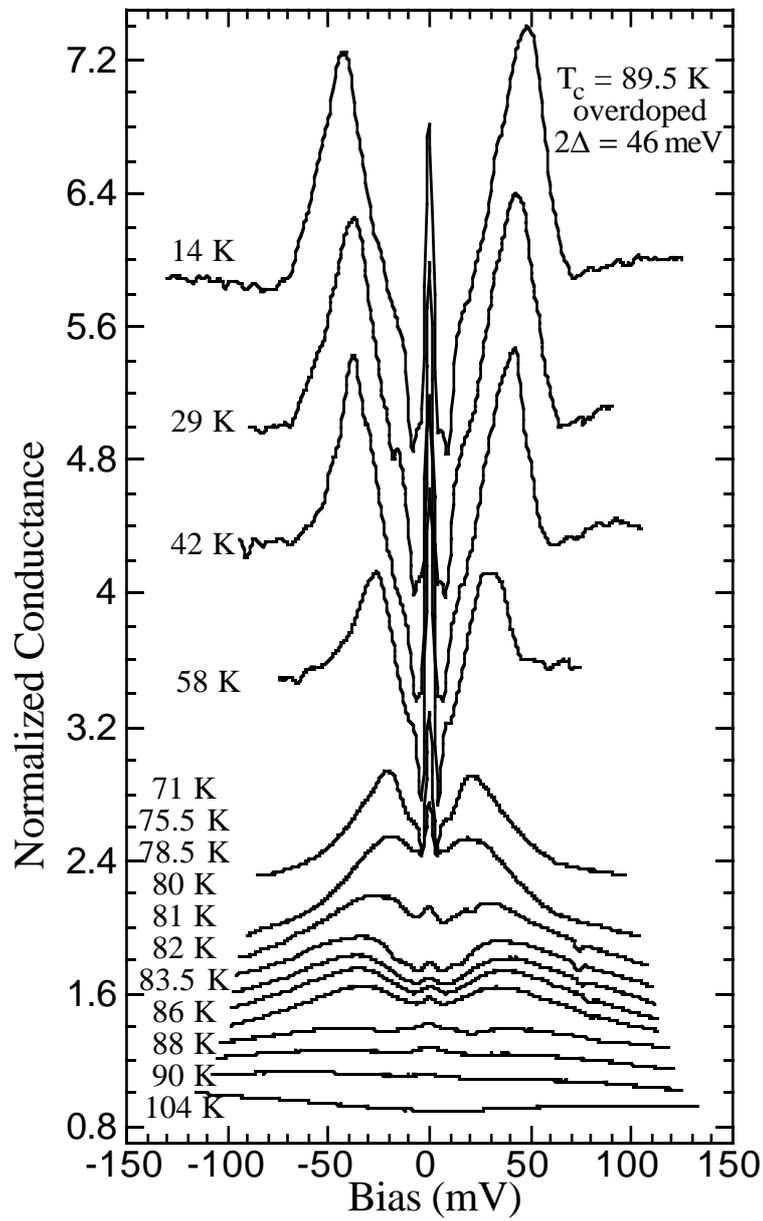

FIG. 14

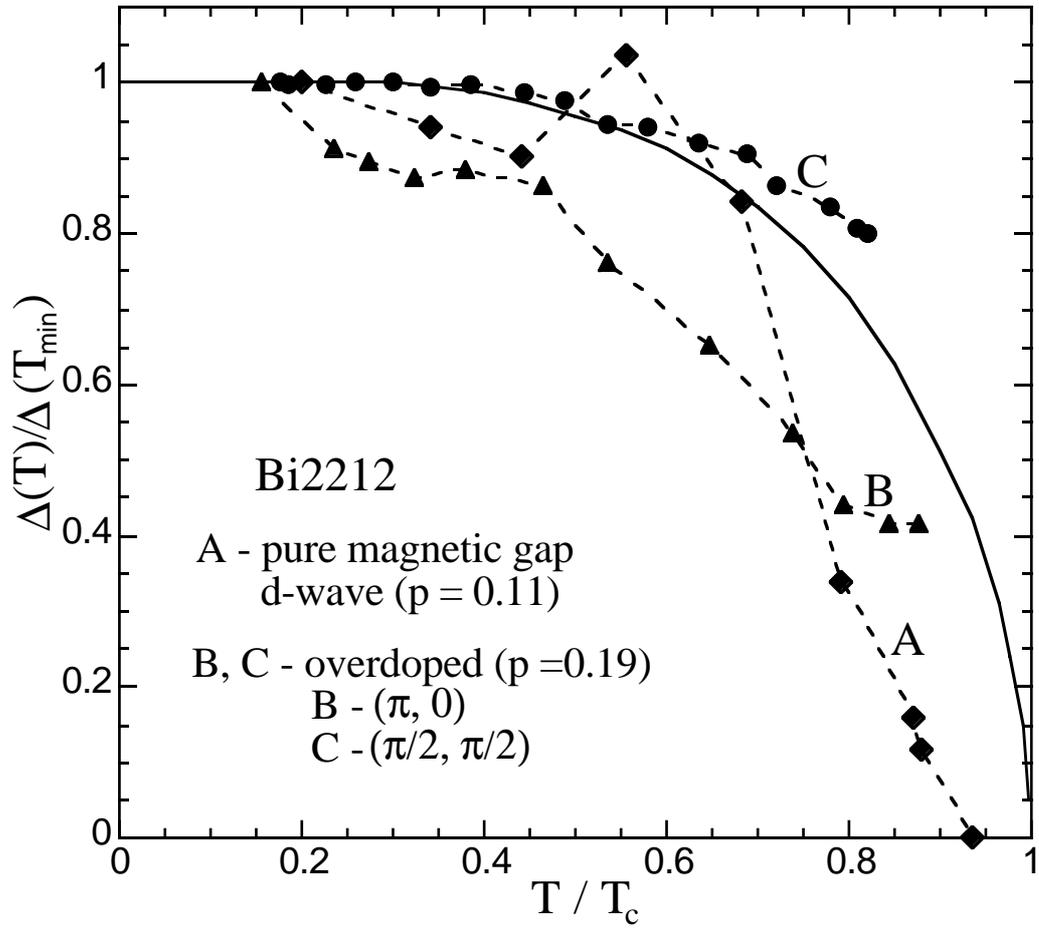

FIG. 15

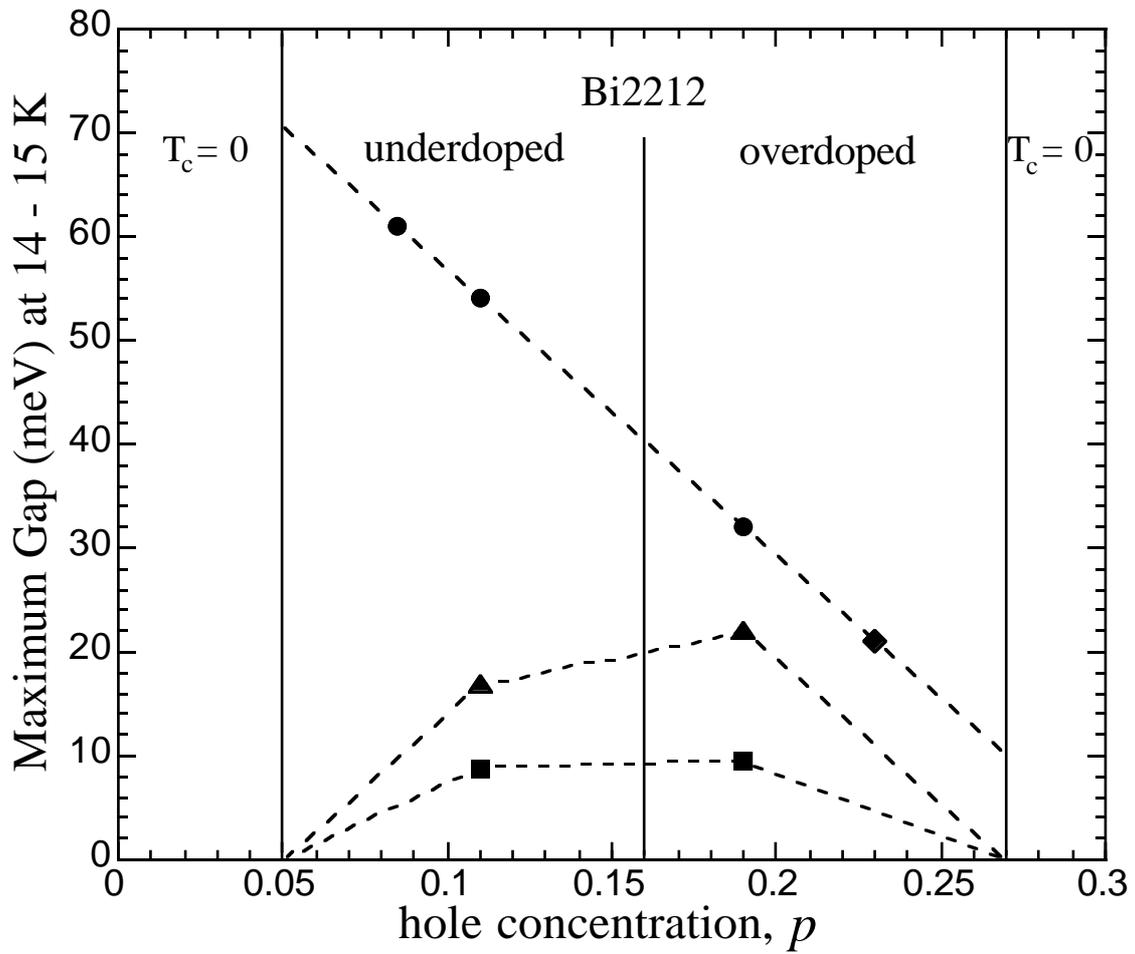

FIG. 16